\documentclass[12pt]{iopart}
\usepackage{cite}					% better citing: hyphen in list of consecutive cites

\usepackage{graphicx}

\newcommand{\subfigimg}[3][,]{%
  \setbox1=\hbox{\includegraphics[#1]{#3}}% Store image in box
  \leavevmode\rlap{\usebox1}% Print image
  \rlap{\hspace*{0pt}\raisebox{\dimexpr\ht1-0.8\baselineskip}{\large#2}}% Print label
  \phantom{\usebox1}% Insert appropriate spacing
}

\begin{document}

\title[]{%
Spatial Confinement Causes Lifetime Enhancement and Expansion of Vortex Rings with Positive Filament Tension
}

\author{Jan Frederik Totz$^1$, 
Harald Engel$^1$,
and Oliver Steinbock$^2$}

\address{$^1$ Institut f\"ur Theoretische Physik, EW 7-1, TU Berlin, Hardenbergstr. 36, 10623 Berlin, Germany}
\address{$^2$ Department of Chemistry and Biochemistry, Florida State University, Tallahassee, Florida 32306-4390, USA}
\ead{jantotz@itp.tu-berlin.de}

\begin{abstract}
We study the impact of spatial confinement on the dynamics of three-dimensional (3d) excitation vortices with circular filaments. In a chemically active medium we observe a decreased contraction rate of such scroll rings and even expanding ones, despite their positive filament tension. All experimentally observed regimes of spatially confined scroll ring evolution are reproduced by full 3d numerical integration of the underlying reaction-diffusion equations. Additionally, we propose a kinematical model that takes into account the interaction of the scroll ring with a no-flux boundary. Its predictions agree quantitatively with data obtained from simulations of the reaction-diffusion model.

%We propose a kinematical model which takes into account the interaction of the scroll ring with a confining Neumann boundary. The model reproduces all experimentally observed regimes of ring evolution, and correctly predicts the results obtained by direct numerical integration of the underlying reaction-diffusion equations.
\end{abstract}

%05.45.-a 	Nonlinear dynamics and chaos (see also section 45 Classical mechanics of discrete systems; for chaos in fluid dynamics, see 47.52.+j)
%82.40.Ck 	Pattern formation in reactions with diffusion, flow and heat transfer (see also 47.54.-r Pattern selection; pattern formation and 47.32.C- Vortex dynamics in fluid dynamics)
%82.40.Qt 	Complex chemical systems (see also 82.39.Rt Reactions in complex biological systems and 87.18.-h Biological complexity)
\vspace{2pc}

\submitto{\NJP}
%\maketitle

\maketitle

%\tableofcontents

\section{Introduction} \label{intro}

Confinement effects attract interest across many areas of physics as they generate a wealth of non-intuitive phenomena. One widely studied example is driven Brownian motion in spatial confinement, which exhibits intriguing features such as suppressed diffusion resulting in a violation of the Einstein fluctuation-dissipation relation \cite{martens_entropic_2011} and hydrodynamically enforced entropic trapping of Brownian particles \cite{martens_hydrodynamically_2013,Keyser2014-1,Keyser2014-2}.
Phase separation in porous materials leads to layering, freezing, wetting and other phase transitions not found in the bulk system, given that the pore size is on the order of the range of the forces between the confined molecules \cite{gelb_phase_1999}.
%
%Widely studied examples are (dynamic processes near plane walls that can affect the speed of human swimmers \cite{swim}) 
Further examples include the efficacy of insect flight \cite{rayner1991vortex}, vortex-related phenomena such as the onset of turbulence \cite{doligalski1994vortex} and many others. \\ \indent
Vortex structures exist also in excitable systems including chemical reaction-diffusion media and information relaying biological systems \cite{agladze2000waves,bansagi2006nucleation,kastberger2008social}.  Confinement effects in excitable media gained interest, because important examples, such as widely studied multicellular organisms \cite{steinbock1993three,zykov2011selection} and the human heart \cite{cherry2011effects}, measure at most a few vortex wavelengths.
Vortices in two-dimensional excitable systems are rotating spiral waves and exist only above a minimal system size \cite{Hartmann1996,yang2006investigation}. Their interaction with a no-flux (Neumann) boundary induces a drift of the spiral tip along the wall in which the tip-wall distance is a system-specific value below the pattern wavelength \cite{brandtstadter2000experimental,schebesch1990,zemlin2005}. 
%
%It has been shown that the effect of a no-flux boundary is formally identical to that of a mirror image \cite{schebesch1990,zemlin2005}. 
%
% It is also known that sharp corners in the system boundary can nucleate spiral waves from non-rotating wave trains \cite{agladze1994rotating}. 
%Moreover, local heterogeneities such as unexcitable discs can pin spiral tips, which then orbit along the boundary of the inclusion at an altered rotation frequency \cite{steinbock1992chemical, Zykov2010797}. \\ \indent
%
%Scroll waves, the 3d generalizations of spiral waves,
3d generalizations of rotating excitation waves were discovered by A. Winfree \cite{Winfree1973}, which he termed scroll waves.
They organize around a one-dimensional phase singularity called filament.
%In 3d media, excitation vortices are called scroll waves and rotate around one-dimensional phase singularities
In the limit of small curvature and twist, the local speed of these filaments is proportional to their local curvature \cite{biktashev1994tension, keener1992dynamics,Panfilov1986, panfilov1987two}. In the case of a positive filament tension $\alpha$, the circular filament of a scroll ring contracts according to a simple square root law \cite{keener1992dynamics}. The vortex annihilates in a finite time $R_0^2/(2\alpha)$ where $R_0$ is the initial filament radius.
 
 For negative $\alpha$, scroll waves undergo negative line tension instability that eventually result in a spatio-temporally irregular regime called vortex or Winfree turbulence \cite{biktashev1994tension, bansagi2007negative}. This case will not be considered here. We emphasize, however, that confinement can suppress the negative line tension instability of a scroll ring and give rise to the formation of an autonomous boundary-stabilized 3d pacemaker \cite{Azhand2014}.
 
\begin{figure}[!ht]
\flushright
\includegraphics[width=0.85\linewidth]{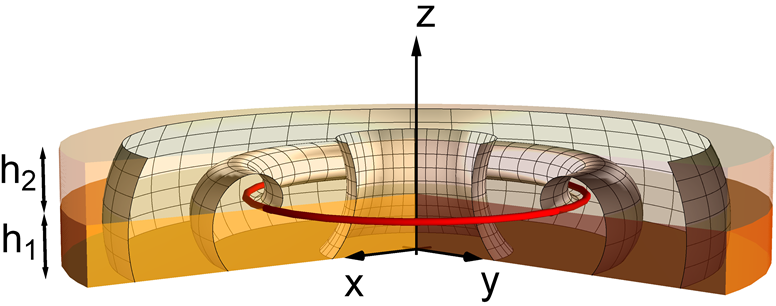}
\caption{\label{fig:exsetup}
Schematic view of the scroll ring (filament shown in red) confined to a cylindrical disk consisting of a gel bottom layer (height $h_1$) covered by a liquid top layer (depth $h_2$). 
}
\end{figure}

Experimental and numerical studies on the formation and evolution of free scroll waves and rings have been reported for example in \cite{Storb2003, Alonso2003, Alonso2006,bansagi2006nucleation,Luengviriya2008, Dahmlow2013}.
%Furthermore waves pinned to localized heterogeneities and their detachment therefrom were investigated \cite{}.
%, because of their relevance to cardiac arrest scenario. 
%
%In addition, Bray and Wikswo simulated "quatrefoil reentry", which involves the interaction of two scroll rings having opposite chirality \cite{bray2003interaction}. This scenario shares similarities to vortex interaction with no-flux boundaries.
%
%A numerical study by Winfree noted the stabilizing case of boundary interaction, but only for negative filament tension \cite{winfree1989}.
%
%However, the influence of planar no-flux boundaries close to rings with positive filament tension has never been studied experimentally.
However, the behavior of vortices with positive filament tension near planar no-flux boundaries has never been studied experimentally. To this end we study the evolution of scroll rings in thin layers of the ferroin-catalyzed Belousov-Zhabotinsky (BZ) reaction (figure \ref{fig:exsetup}), \cite{Linde1991, amemiya_formation_1998, bansagi2006nucleation, Jimenez2012}.
%
%figure_1_experimental_images2.png
%------------------------------------------ fig1 ---------------------------------------------%
%----------------------------------------- end fig1 ---------------------------------------------%
%
%We demonstrate experimentally that the interaction of the wave fronts with the no-flux top and bottom boundaries of the layer can significantly alter the ring dynamics as compared to a scroll ring in an unbounded medium. A kinematical model leads to good agreement between measurements and numerical simulations based on a reaction diffusion model of the BZ reaction \cite{Rov1984, Aliev1992}.
We demonstrate experimental evidence that spatial confinement leads to qualitatively new ring dynamics unobtainable in an unbounded medium (chapter \ref{experiments}). The experimental observations are reproduced in full 3D numerical simulations of the underlying  Rovinsky-Aliev reaction-diffusion model \cite{Rov1984, Aliev1992} (chapter \ref{numerics}). 
%Finally we propose a kinematical model that agrees quantitatively with the simulated scroll ring trajectories (chapter 4). 

Finally we propose a kinematic approach that takes into account the interaction of the scroll ring with a no-flux boundary. The predictions of the kinematical model for the ring dynamics in a phase plane spanned by the filament radius and the distance between filament plane and boundary are in quantitative agreement with data obtained from 3d numerical simulations of the reaction-diffusion model (chapter \ref{kinematics}).

%The predictions (about the ring dynamics in a phase plane spanned by the filament radius and the distance between filament plane and boundary are in quantitative agreement with data obtained from 3d numerical simulations of the reaction-diffusion model.
% ?

\section{Experiments} \label{experiments}

%figure_1_experimental_images2.png
%------------------------------------------ fig1 ---------------------------------------------%
\begin{figure}[!ht]
\flushright
\includegraphics[width=0.85\linewidth]{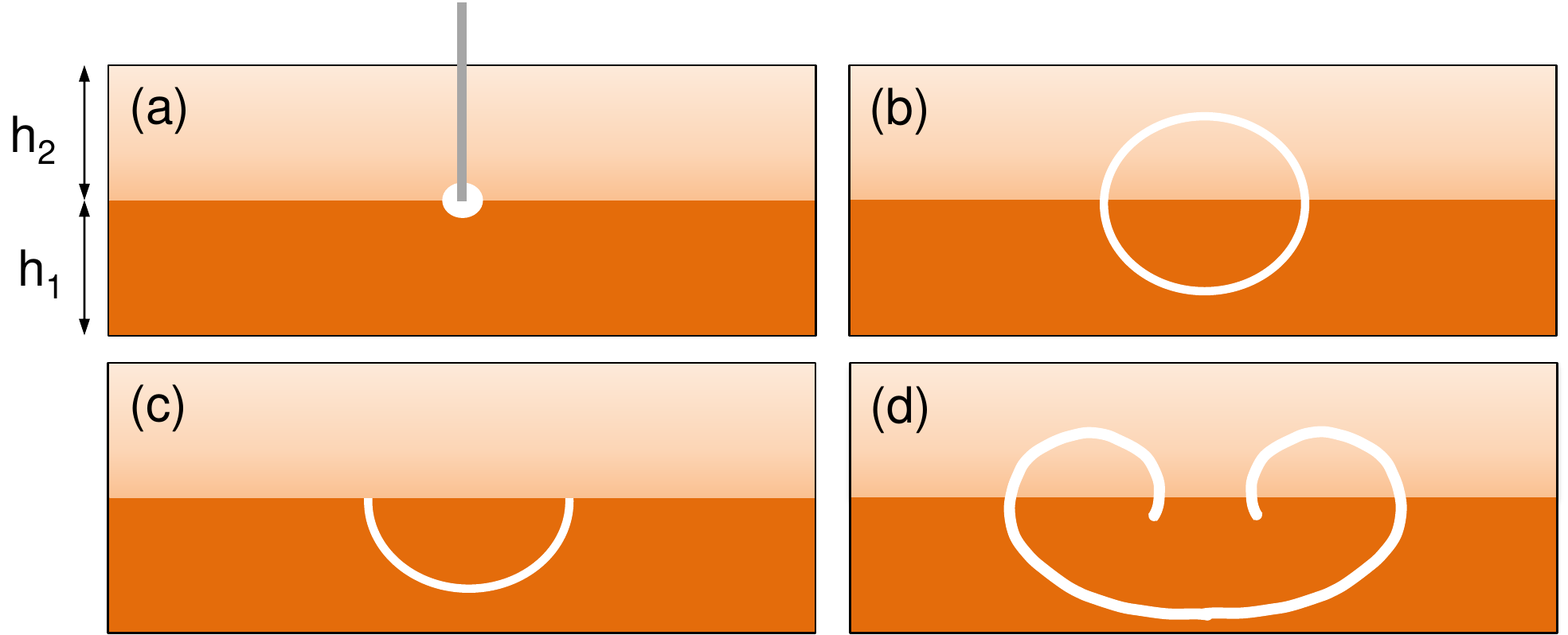}
\caption{\label{fig:exsetup2}
Illustration of scroll ring initiation in the bipartite medium. (a) A silver wire induces a spherical wave, after which the wire is removed. (b,c) As soon as the nucleated wave has grown to a certain size, the medium is shaken strongly, erasing the wave in the upper liquid part (bright).  (d) The open ends grow out of the lower gel part (dark), curl in and form a planar scroll ring.
}
\end{figure}
%---------------------- end fig1 ---------------------------------------------%

%figure_1_experimental_images2.png
%------------------------------------------ fig1 ---------------------------------------------%
\begin{figure}[!ht]
%\flushright
\includegraphics[width=1\linewidth]{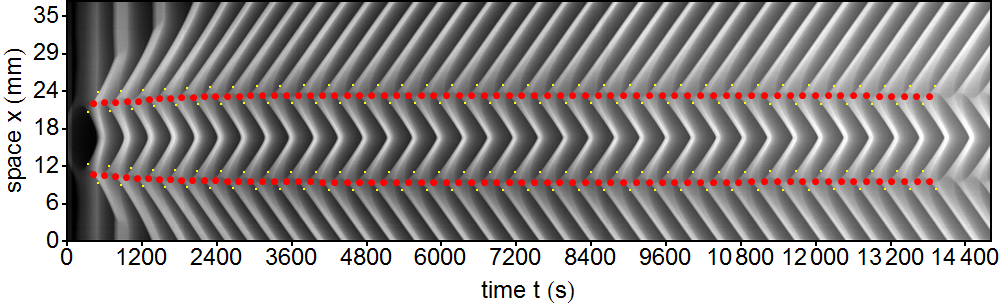}
\caption{\label{fig:auto}
Computer-assisted localization of the filament loop. Each outward and inward traveling wave is identified and fitted with a linear function. The intersection points (yellow) of extrapolated subsequent fits allow for determination of the filament location (red dots here and red lines in figure \ref{fig:exresults}). In this example the contraction rate of the filament is vanishingly small.
}
\end{figure}
%----------------------------------------- end fig1 ---------------------------------------------%

%%%%%%%%%%%%%%%%%%%%%%%%%%%%%%%%%%%%%%%%%%%%%%%%%%%%%%%%%%%%%%%%%%%%%%% Experimental Results %%%%%%%%%%%%%%%%%%%%%%%%%%%%%%%%%%%%%%%%%%%%%%%%%%%%%%%%%%%%%%%
%CH2COOH2
The reaction takes place in a sealed Petri dish (figure \ref{fig:exsetup2}). This measure prevents external oxygen from affecting the reaction \cite{taylor_scroll_1999}.
Throughout all experiments the same initial reactant concentrations are used: $\mathrm{[H_2SO_4]=0.16\,M}$, $\mathrm{[NaBrO_3]=0.04\,M}$, $\mathrm{[malonic\;acid]=0.04\,M}$ and $\mathrm{[Fe(phen)_3SO_4]=0.5\,mM}$. 
The BZ medium consists of a bottom gel layer (agarose 0.8\,\% weight/volume) and a top liquid layer. 
Both layers are prepared from the same concentrations.
For these conditions the filament tension of an unbounded scroll ring was previously determined to be as $\mathrm{\alpha=1.4\times 10^{-5}\:cm^2\,s^{-1}}$.
Curvature-induced filament motion in binormal direction was shown to be vanishingly small \cite{Jimenez2012}.
%
% Boundary conditions
% No-flux boundary conditions because it is bounded by the extent of the Petri dish and  another glass plate on top
%
The total medium height is varied in different experiments between $\mathrm{5.2-8.0\,mm}$. The spatial extension in the horizontal direction is bounded by the Petri dish diameter ($\mathrm{90\,mm}$).
Top and bottom boundaries are planar, parallel plexiglass surfaces.

To initiate the scroll ring, we introduce a thin wire of pure silver (99.9\,\%) at the interface of the two layers for about $\mathrm{20\,s}$ (figure \ref{fig:exsetup2}a). This decreases the local concentration of inhibitory bromide ions in the direct vicinity of the wire tip. 
As soon as the developing spherical wave reaches a certain size (figure \ref{fig:exsetup2}b), the system is strongly agitated to spatially homogenize the liquid top layer (figure \ref{fig:exsetup2}c). After the fluid comes to rest again, the unperturbed gel-bound part of the wave extends into it and starts to curl in, thereby nucleating the scroll ring as depicted in figure \ref{fig:exsetup2}d).
All scroll rings are initialized in sufficient distance ($\mathrm{>30\,mm}$) from the lateral boundaries to exclude their influence.

A charged coupled device camera mounted over the system records transmission image sequences. Grayscale values in the recorded images are proportional to the light transmission integrated across the total height of the active medium \cite{Muller1987}. The faint contrast between the blue wave and the red refractory wake (corresponding to the oxidized and reduced state of the catalyst, respectively) is enhanced using an additive dichroic filter.

\begin{figure}[!t]
\flushright
\includegraphics[width=0.98\linewidth]{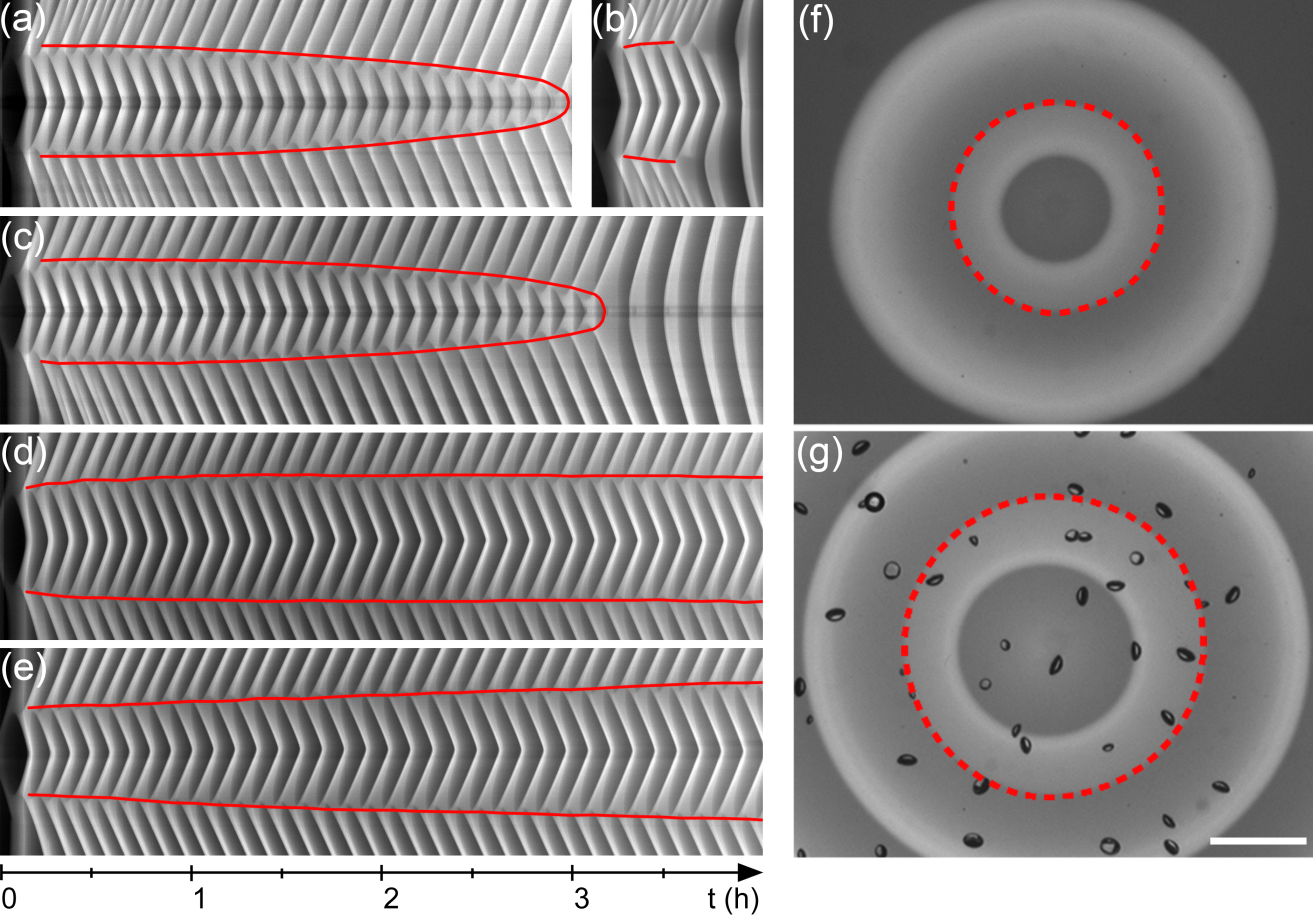}
\caption{\label{fig:exresults}
Experimentally observed scenarios for the evolution of the scroll ring.
Panels (a) -– (e) display space-time plots extracted from a sequence of recorded transmission images with the filament position overlaid in red. The vertical space axis and the horizontal time axis span $\mathrm{2.3\,cm}$ and
(a) $\mathrm{3\,h}$, (b) $\mathrm{50\,min}$, (c--e) $\mathrm{4\,h}$.
The plots are obtained from experiments with $h_1$ and $h_2$ equal $\mathrm{4\,mm}$ and $\mathrm{4\,mm}$ (a), $\mathrm{3\,mm}$ and $\mathrm{4\,mm}$ (b), $\mathrm{4\,mm}$ and $\mathrm{3\,mm}$ (c), $\mathrm{4\,mm}$ and $\mathrm{2.5\,mm}$ (d), $\mathrm{4\,mm}$ and $\mathrm{1.25\,mm}$ (e, f, g), respectively. (f) and (g): Two snapshots demonstrating the expansion of the filament loop over the course of 3 hours. The white bar corresponds to $\mathrm{5\,mm}$.
}
\end{figure}

\begin{figure}[!ht]
\flushright
\includegraphics[width=0.75\linewidth]{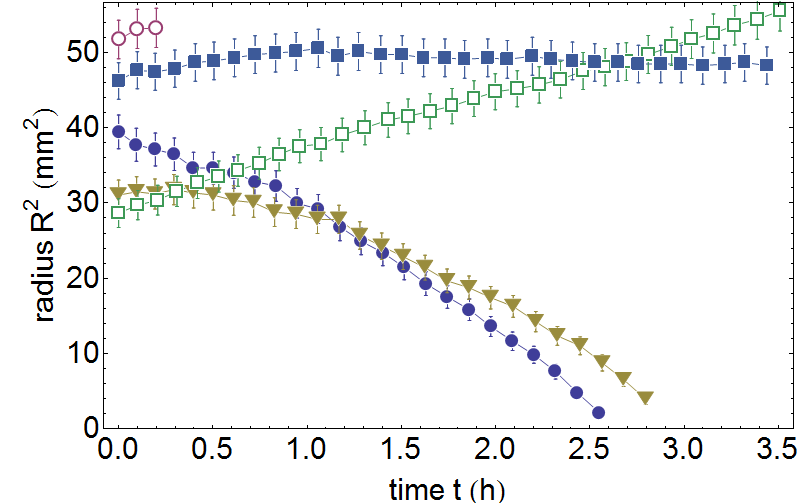}
\caption{\label{fig:exresultsr2}
Squared filament radius plotted versus time. The data are obtained from space-time plots in figure \ref{fig:exresults}.
Open circles: Annihilation of the scroll ring (figure \ref{fig:exsetup2}b) after collision with the bottom boundary.
%The R versus time dependence coincides with that of an unconfined scroll ring. 
Full circles: Contracting free scroll ring (figure \ref{fig:exsetup2}a) outside the interaction range with the layer boundaries. Note, that $R^2$ decreases almost linearly with time.
Triangles: Collapsing scroll ring (figure \ref{fig:exsetup2}c) with significantly increased life time as compared to (figure \ref{fig:exsetup2}a).
Full squares: Scroll ring (figure \ref{fig:exsetup2}d) contracting at a vanishingly small rate.
Open squares: Expanding scroll ring (figure \ref{fig:exsetup2}e).
Error bars result from the spatial ambiguity in the filament location.
}
\end{figure}

After the experiment, the series of transmission images is reduced to a space time plot by extracting pixels along a line that passes through the center of the circular scroll ring. We reconstruct the filament location from the plots based on a method developed in \cite{pertsov_three-dimensional_1993}
. Each wavefront originating from the filament is recognized by image processing and fitted by a linear function. The intersection points of extrapolated fits from a pair of subsequent outward and inward traveling waves allow for a localization of the filament (figure \ref{fig:exresults}). From these data, we can extract the time evolution of the filament radius (figure \ref{fig:exresultsr2}) and the life time of the scroll ring.

%figure_1_experimental_images2.png
%------------------------------------------ fig1 ---------------------------------------------%
%----------------------------------------- end fig1 ---------------------------------------------

%Our experiments show that contraction can be strongly delayed (Fig. \ref{fig:exsetup}f), resulting in quasi-stationary, almost `immortal' scroll rings (Fig. \ref{fig:exsetup}g). 
Our experiments reveal that the close vicinity of a Neumann boundary can strongly delay the contraction of the filament loop. The scroll ring on figure \ref{fig:exresults}c lives longer than the ring shown in figure \ref{fig:exresults}a even though its initial radius is smaller. While the larger ring (full circles in figure \ref{fig:exresultsr2}) follows the known square root contraction for the radius \cite{keener1992dynamics}, the more enduring scroll ring (triangles) decreases slowly in the early stages. Even though the later stages feature a faster decrease, the contraction rate is still below its larger counterpart. This example clearly violates the monotonic lifetime dependency on $R_0$.

Under stronger spatial confinement we even observe persistent, almost stationary scroll rings (figure \ref{fig:exresults}d). In the beginning of the experiment, the ring slightly grows in size until it reaches a maximum after about an hour. Then a nearly vanishing contraction ensues. 
%with an effective filament tension one order of magnitude smaller than what is measured before in the system ($\mathrm{\alpha=1.2 \times 10^{-6}\,cm^2s^{-1}}$). 
Over the course of the experiment, the ring lasted for 51 periods and did not collapse.

Moreover, despite of positive filament tension, ring contraction can be superseded by expansion (figure \ref{fig:exresults}e). For this example, we choose a small ring in a very thin medium of height 5.25\,mm. In contrast to all other cases, this ring expands during the complete experiment and reaches a larger size than of any other ring. Notice the nearly linear increase of $R^2$ (open squares) for the experimental conditions also shown in figure \ref{fig:exresults}e. The corresponding increase in $R$ equals about $40\%$ over 3.5 hours (figure \ref{fig:exresults}f,g). 
Eventually the ring breaks apart due to interaction with $\mathrm{CO_2}$ bubbles. These localized inhomogeneities are a product of the BZ reaction and grow in number as well as size over the course of an experiment.
 
%---------------------------------------------------------- fig2 ---------------------------------------------%
%---------------------------------------------------------- end fig2 ---------------------------------------------%

Conversely, the life time of the scroll ring can be drastically reduced by sudden annihilation of the filament with a nearby boundary (figure \ref{fig:exresults}b). Instead of an expected lifetime of $3.5$\,hours the ring disappears after $15$\,minutes. %Interestingly the complete filament does not vanish simultaneously. First, only a small arc disappears and is then followed by the remainder a period later. This indicates that the filament plane was not perfectly parallel to the boundary at the end.

\newpage
\section{Numerical simulations based on the Aliev-Rovinsky model} 
\label{numerics}

Our experimental results suggest that the interaction of the scroll ring with a confining no-flux boundary is responsible for the observed modifications in the scroll ring dynamics as compared to the spatially unbounded case.
In support of this hypothesis, now we study the interaction of a planar scroll ring with a no-flux boundary numerically. To this end we solve the two-component reaction-diffusion model that was developed by Aliev and Rovinsky especially for the ferroin-catalyzed BZ reaction based on the FKN-mechanism \cite{Field1972}. The model is given by the equations

%Our experimental setup limits us to the observation of radius dynamics. The height coordinate of the filament plane and its distance to the boundary can not be determined. A tomographic reconstruction is not possible due to light absorption in the medium. Since we require a diameter that is large enough to neglect interactions with the radial boundaries, the measurable light intensity in radial direction is too weak. 

%Numerical simulations are free from these limitations and allow us to accurately track the radius of the filament ring as well as the distance to the boundary. We choose an Oregonator-class model \cite{Aliev1992} developed specifically for the ferroin-catalyzed BZ reaction:

\begin{equation} %\nonumber
\label{rds}
\eqalign{
	\frac{\partial u}{\partial \tau}  =  \frac{1}{\epsilon}\left[u-u^{2}-\left(2qa\frac{v}{1-v}+b\right)\frac{u-\mu}{u+\mu}\right]+D_{u}\nabla^2 u, \cr
	\frac{\partial v}{\partial \tau}  =  u-a\frac{v}{1-v}+D_{v}\nabla^2 v.
	}
\end{equation}
Here $u$ and $v$ are proportional to the concentrations of bromous acid and ferriin, respectively. 
We calculated the values for the parameters $a, b, \epsilon, q$ and  $\mu $  from our recipe concentrations, diffusion coefficients $D_u$ and $D_v$ are the same as in \cite{bray2003interaction,Aliev1992}.
%
%From simulation, we obtain a value of $\lambda_{sim} = 0.56\,\mathrm{cm}$ as the wavelength and $T_{sim} = 372\,\mathrm{s}$ as the rotation period of a free spiral wave. 
From 2d simulations of a free spiral wave, we obtain the wavelength $\lambda_{sim} = 0.56\,\mathrm{cm}$ and the rotation period $T_{sim} = 372\,\mathrm{s}$.
The corresponding values in the experiment agree very well: $\lambda_{exp} = 0.58\,\mathrm{cm}$ and $T_{exp} = 390\,\mathrm{s}$.
% 
%In addition we find 

% how do others say it?
%The calculations were performed in a three-
%dimensional domain using a cylindrical coordinate sys-
%tem ?z;?;??.
% Computations
% were performed using finite differences and an explicit
% time integration scheme on grids containing up to 200 ?
% 200 ? 200 elements in 3D and 700 ? 700 in 2D
Our full 3d simulations are performed in Cartesian coordinates. In contrast to the computationally less demanding use of cylindrical coordinates, this allows for potential angular instabilities to develop along the ring. Lateral boundaries were at least two wavelengths away from the filament to ensure that their influence were negligible. Furthermore we used parallelized forward Euler integration with a time step of $\mathrm{d\tau=0.05}$ and a 7-point Laplacian stencil with $\mathrm{dx=0.001}$. The filament is determined as the set of points which form the intersection of two level sets of the activator variable $u$ from subsequent time steps. 
% Viel mehr details! im unbounded: free is found, Free, R^2 (t), alpha bestimmt aus slope

%In our simulations we use the same set up as in the experiment (compare figure 1). 
The initiation of the scroll ring mimics the experimental procedure described in figure \ref{fig:exsetup2}. The scroll rings interact with the bottom Neumann-boundary.
%The integration area is chosen so large, that the interaction of the filament with the lateral boundaries is excluded.
%
%Simulations show damped contraction, as observed in the experiment, only in close distance to the bottom noflux boundary. The top boundary accelerates the contraction for the given scroll ring chirality. In the following we will focus on the influence of the bottom boundary alone for the sake of clarity. We define our simulation medium with a height of two wavelengths. Simulations in a narrower system show the same qualitative solutions.
In simulations with a large cubical grid of edge length $6\,\lambda_{sim}$ we find unperturbed scroll rings contracting according to the well-known square root dependence $R(t)=\sqrt{R_0^2-2\alpha t}$ predicted by Keener et al. \cite{keener1992dynamics}. From a linear fit of $R^2$ versus time we obtain for the filament tension $\alpha_{sim}=0.024\,\lambda_{sim}^2\,T_{sim}^{-1}$, which is in very good agreement with the experimentally measured value $\alpha_{exp}=0.026\,\lambda_{exp}^2\,T_{exp}^{-1}$. 

%For the filament tension in the unbounded case, and 
% erkläre Probleme mit dominating wall

%------------------------ fig6 ---------------------------------------------%
\begin{figure}[!ht]
	\flushright
	\begin{tabular}{@{}p{0.96\linewidth}@{\quad}p{\linewidth}@{}}
		\flushright 
		\subfigimg[width=0.95\linewidth]{\hspace{-0.3cm}(a)}{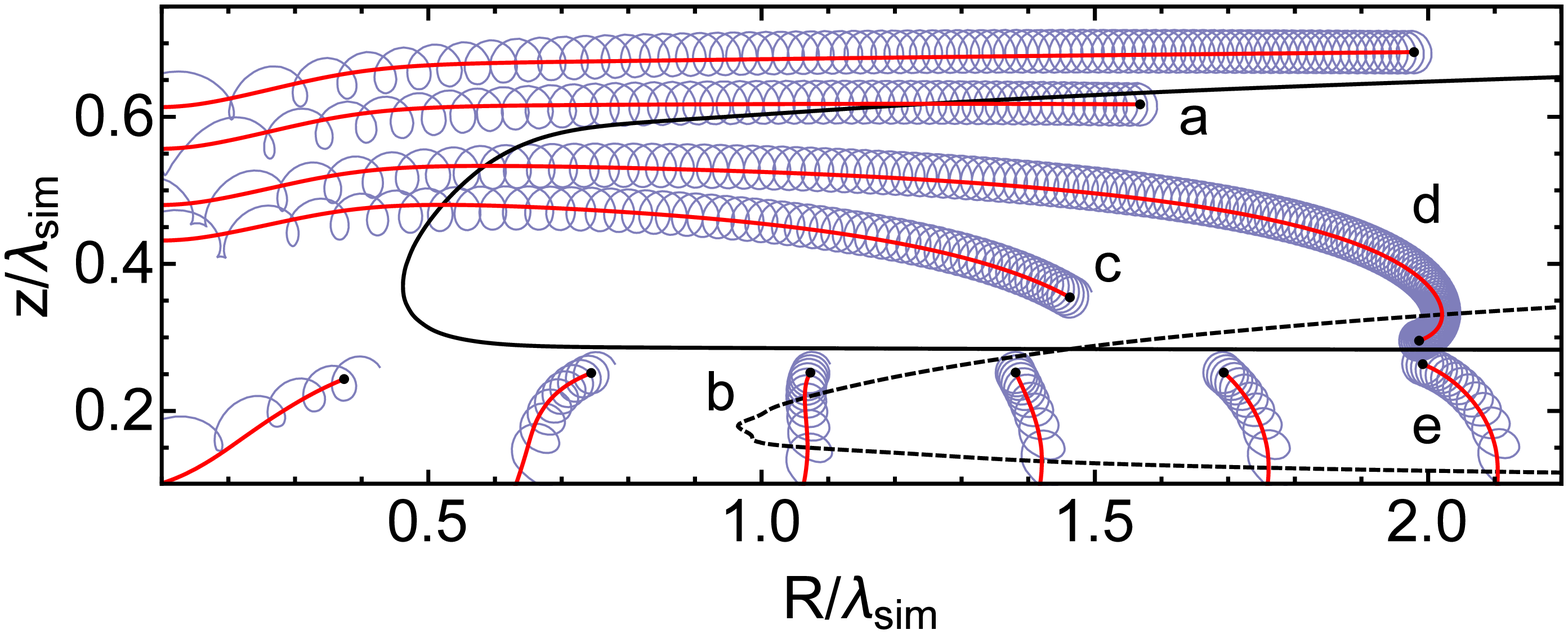} \\
		\vspace{0.4cm}
		\subfigimg[width=0.95\linewidth]{\hspace{-0.3cm}(b)}{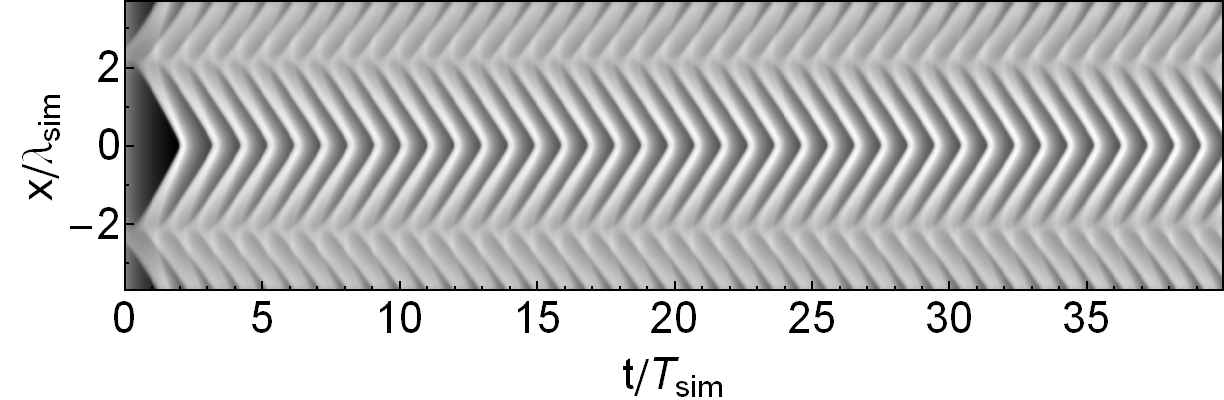}
	\end{tabular}
	\caption{\label{fig:numtrajs}
		(a) Numerically calculated evolution of scroll rings interacting with a no-flux boundary at $z = 0$ in the parameter plane spanned by the distance from the boundary $z$ and the filament radius $R$. Different initial conditions are labelled by black dots. The black dotted and black continuous curves are the nullclines $\dot{R}=0$ and $\dot{z}=0$ of (3), respectively. Letters (a) to (e) correspond to experimentally observed regimes of ring evolution shown in figure \ref{fig:exsetup}. The trajectories of the filament center (thin red lines) calculated from the extended kinematical model (chapter \ref{kinematics} below) are in quantitative agreement with 3d numerical simulations of the BZ model (thin blue lines). (b) Space-time plot of the persistent ring of case d. Even though it is a transient, the ring configuration appears stationary during the initial 40 periods. Parameter values of the BZ model \eref{rds}: $\epsilon=0.47$, $a=0.018$, $b=0.00039$, $\mu=0.00051$, $q=0.5$, and $D_u=D_v=2\times10^{-5}\:\mathrm{cm^2\,s^{-1}}$. Kinematical parameters (3): $\alpha=0.024$ and $\beta=-0.0004$. 
	}
\end{figure}
%--------------------------------------- end fig7 ------------------------------------%

\newpage
The results of our numerical simulations of the Aliev-Rovinsky equations are summarized in figure \ref{fig:numtrajs}, where we have plotted trajectories in the R-z-plane with $R(t)$ and $z(t)$ denoting the radius of the filament and the distance of the filament plane from the confining boundary, respectively. Depending on the initial conditions $R_0 = R(t=0)$ and $z_0 = z(t=0)$ we find different scenarios for the evolution of the scroll ring. Solutions that show remarkable similarity to regimes observed experimentally are labelled (a) to (e) to facilitate the comparison between experimental (figure \ref{fig:exresults}) and numerical (figure \ref{fig:numtrajs}) results.
%In our simulations (figure \ref{fig:numtrajs}), we can identify sets of solutions  that show remarkable similarity to our experimental findings. We highlight this relationship by using the same letter for the numerical solutions in figure \ref{fig:numtrajs} as for the subfigures of figure \ref{fig:exresults}.

\section{Kinematical approach} \label{kinematics}

In an unbounded medium scroll rings are instationary objects. Their radius, $R(t)$, shrinks or expands with time depending on whether the filament tension $\alpha$ is positive or negative, respectively. Simultaneously, an axisymmetric scroll ring in general drifts along its symmetry axis, $z(t)$. For not too small filament curvature $1/R$ the time evolution of $R$ and $z$ is given by \cite{panfilov1987two,keener1992dynamics,biktashev1994tension}
\begin{equation}
\label{kinmodel}
\eqalign{
dR/dt = -\alpha/R,  \cr
dz/dt \, = \;\;\;\: \beta/R.
}
\end{equation}

%In the following we argue that the experimentally observed regimes of scroll ring evolution, in particular the life time enhancement, can be explained by the interaction of the ring with a confining no-flux boundary. To this end 

In this chapter we propose an extension of these kinematical equations to account, first, for the interaction of the wave fronts with a confining no-flux boundary and, second, for the intrinsic interaction between wave fronts forming the ring which becomes important at small radii.

In 2d media, the interaction of a spiral wave with a no-flux boundary has been studied extensively both experimentally and theoretically \cite{Ermakova1989,Gomez-Gesteira1996b,brandtstadter2000experimental,Biktasheva2010}. For a spiral wave a no-flux boundary acts effectively as a resonant periodic perturbation that causes a drift of the spiral core provided the ladder is close enough to the boundary. In general the interaction range is smaller than the spiral wavelength, because several wave fronts between spiral core and boundary shield the core region from the perturbing boundary.

\begin{figure}[!ht]
	\flushright
	\begin{tabular}{@{}p{0.80\linewidth}@{\quad}p{\linewidth}@{}}
		\flushright \subfigimg[width=0.85\linewidth]{(a)}{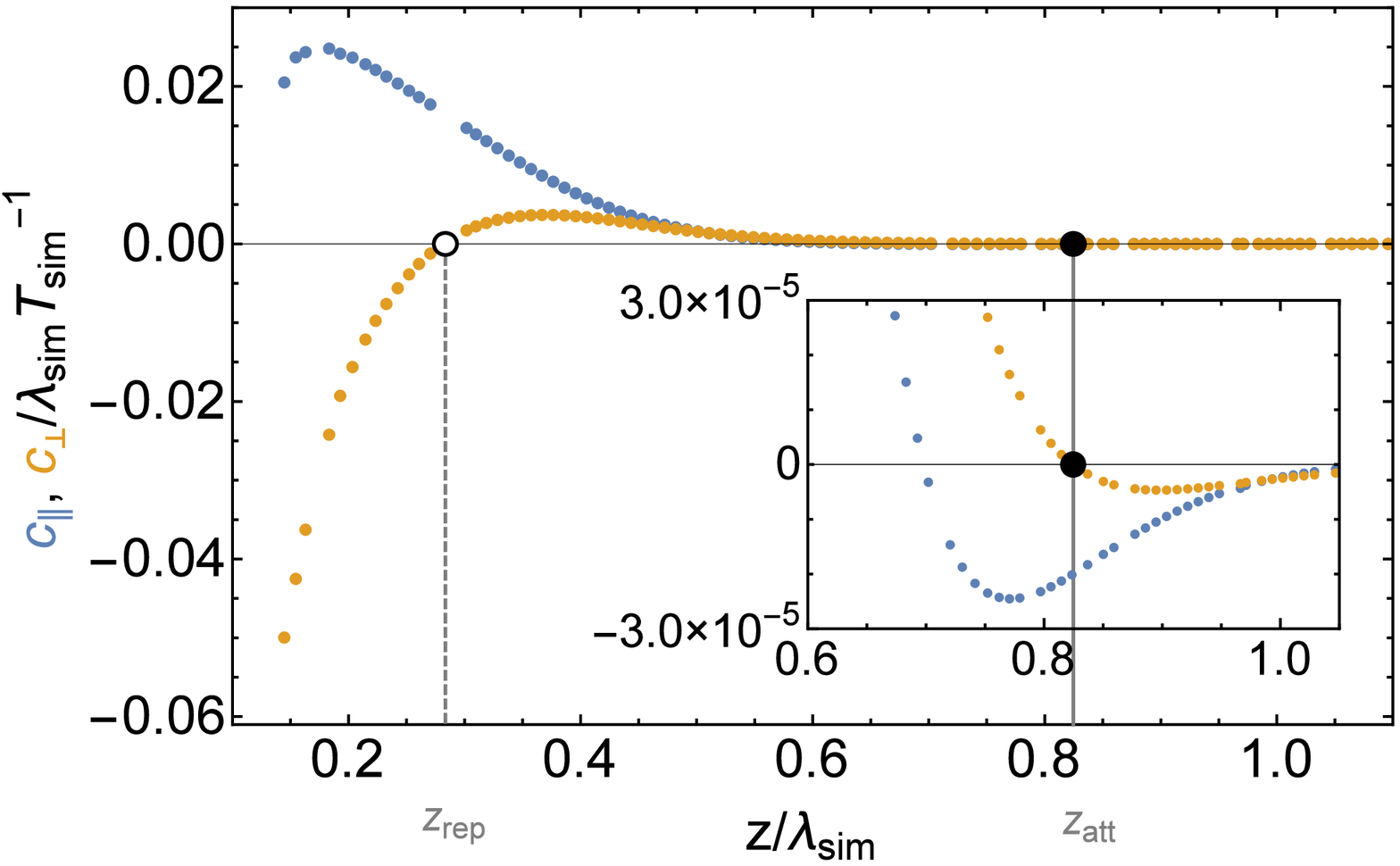} \\
		\subfigimg[width=0.80\linewidth]{\hspace{-0.6cm}(b)}{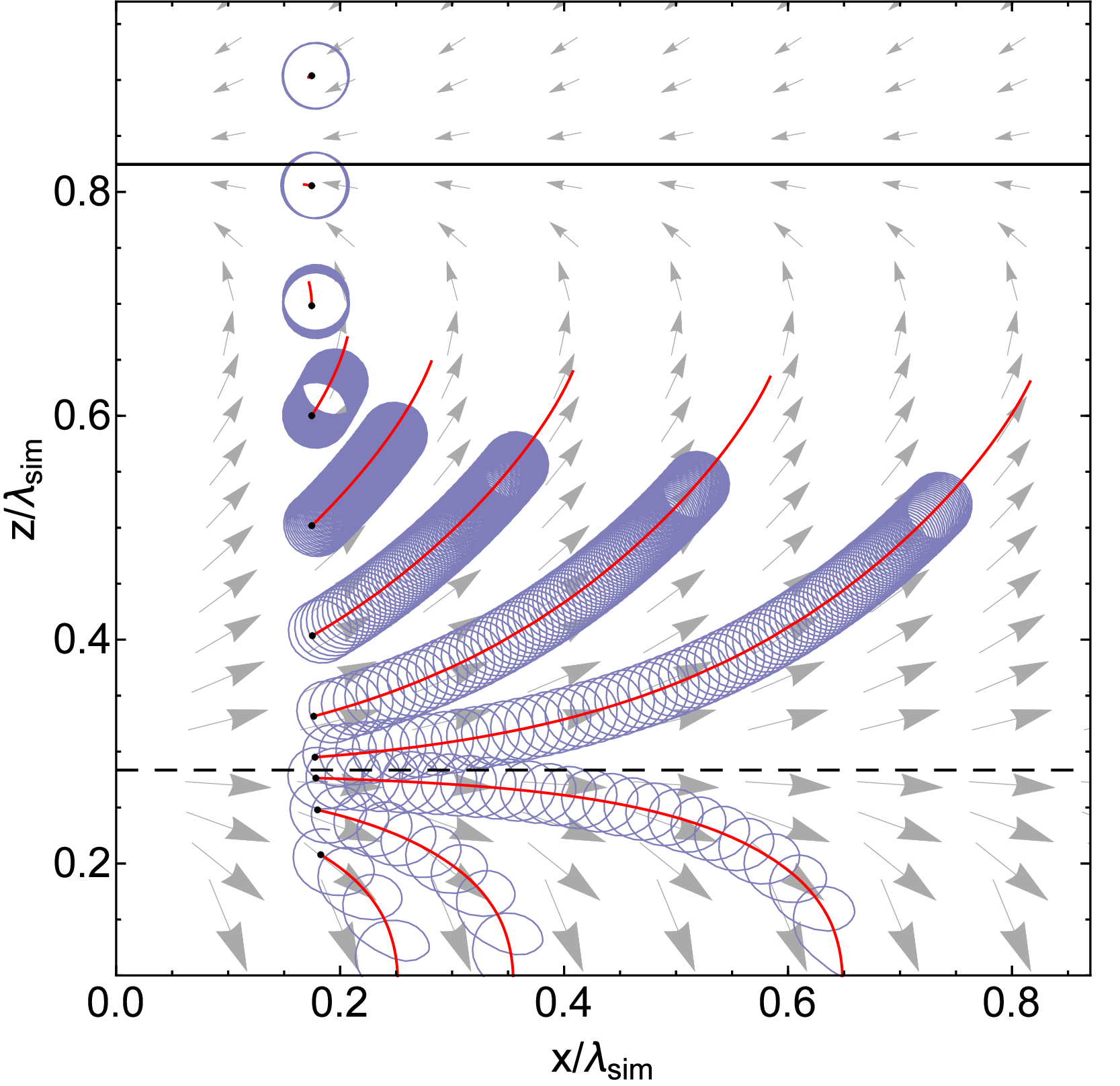}
	\end{tabular}
	\caption{
		Drift of a 2d rigidly rotating spiral wave interacting with a planar no-flux boundary along the x-axis. 
		a) Transversal (orange) and longitudinal (blue) component of the velocity field of boundary-induced spiral drift. Zero-crossings of the transversal drift velocity component at $z =: z_{rep} = 0.28\,\lambda_{sim}$ and $z =: z_{att} = 0.82\,\lambda_{sim}$ (inset) correspond to repelling and attracting distances from the boundary. 
		b) Nonlinearly scaled gray arrows visualize the velocity field of boundary-induced spiral drift $c(z) = ( c_{\parallel}^{2d}(z), c_{\perp}^{2d}(z) )$. The red curves show the motion of the core center obtained by solving the equations $\dot{x}=c_{\parallel}^{2d}(z)$ and $\dot{z}=c_{\perp}^{2d}(z)$. Trajectories of the spiral tip calculated by solving the corresponding BZ model \eref{rds} are shown in blue.  Model parameters are the same as in figure \ref{fig:numtrajs}
		%Drift velocity field of a rigidly rotating spiral wave interacting with a planar no-flux boundary along the x-axis. 
		%a) Perpendicular (red) and parallel (blue) drift velocity components. 
		%All distances are given in units of the spiral wave length $\lambda_{sim}$.
		%All values are given in units of the spiral wavelength $\lambda_{sim}$ and rotation period $T_{sim}$.
		%The first zero-crossing of the normal component at $z=z_{rep}$ corresponds to the unstable drift attractor. Inset: At the distance of its second zero-crossing at $z=z_{att}$, the magnitude of both components is too small to play a dominating role in the dynamics.
		%b) Nonlinearly scaled gray arrows visualize the velocity field. Red curves are solutions of the drift velocity field components $\dot{x}=c_{\parallel}^{2d}(z)$ and $\dot{z}=c_{\perp}^{2d}(z)$, shown in a) and blue curves are spiral tip trajectories of the corresponding 2d reaction diffusion simulation.
		% obtained from averaging the blue spiral tip trajectories. 
		%The full (dashed) horizontal line marks the position of the attracting (repelling) distance at $z_{att}$ and $z_{rep}$, respectively. 
		%
	}
	\label{fig:numdriftfield}
	
\end{figure}

Already in 2d, the dependence of the transversal and longitudinal components of the drift velocity field on the distance of the core centre to the boundary z is difficult to measure experimentally. Therefore, we calculated the components of the drift velocity field from 2d numerical simulations of the Aliev-Rovinsky model of the ferroin catalyzed BZ which was already used for the full 3d simulations. Figure \ref{fig:numdriftfield} displays both drift velocity components $c_{\perp}^{2d}(z)$ and $c_{\parallel}^{2d}(z)$ in units of the wavelength, $\lambda_{sim}$, and rotation period, $T_{sim}$, of the unperturbed spiral wave for different distances $z$ of the filament plane to the no-flux boundary. The normal velocity component $c_{\perp}^{2d}(z)$ has roots at $z = z_{rep} = 0.28\,\lambda_{sim}$ (dashed line in figure \ref{fig:numdriftfield}b) and $z = z_{att} = 0.82\,\lambda_{sim}$. $z_{rep}$ defines a critical distance separating spiral waves that finally annihilate with the boundary from those attracted into a stable regime of constant drift parallel to the boundary at distance $z_{att}$. Note that for the chosen parameters, the drift velocity at distances comparable to $z_{att}$ is already extremely small. For a spiral wave starting shortly above $z_{rep}$ it takes about 10,000 rotation periods to reach $z_{att}$.
This behavior is expected due to the known exponential decay of the interaction strength between a spiral wave and a Neumann boundary \cite{Aranson1991, Aranson1993}.

%-------------------------------------------------------- end fig3 
%We decompose the resulting drift field into a normal and a parallel component relative to the boundary.
%Figure \ref{fig:numdriftfield} displays the dependence of both components $c_{\perp}^{2d}(z)$ and $c_{\parallel}^{2d}(z)$ as a function of the boundary distance and the spatial drift velocity vector field. 
%The continuous functions for the components are high order polynomials fitted to spiral tip trajectories. 

%In three dimensions the two-dimensional boundary induced attractor structure becomes apparent for large scroll rings. However, to understand small scroll rings, it is necessary to take curvature effects into account. 

Based on the 2d simulations, we extend the kinematical model for an unperturbed scroll ring \eref{kinmodel} %\cite{keener1992dynamics} 
by boundary induced drift fields:
\numparts
\begin{eqnarray} 
dR/dt & = & - \alpha/R + c_{\parallel}^{2d}(z) + c_{\perp}^{2d}(R), \label{rmod}
\\
dz/dt & = & \;\;\;\:\beta/R + c_{\perp}^{2d}(z) - c_{\parallel}^{2d}(R). \label{zmod}
\end{eqnarray}
\endnumparts
The first term on the right-hand side of equation \eref{rmod} describes the contraction of the unconfined scroll ring. Functions $c_{\parallel}^{2d}(z)$ and $c_{\perp}^{2d}(z)$ denote the normal and tangential velocity components of the boundary-induced drift of a 2d spiral wave interacting with a planar no-flux boundary discussed above. 
% {\tiny Bray} et al. \cite{bray2003interaction} ...???
%
%
%///// ???
%
Velocity components $c_{\perp}^{2d}(R)$ and $c_{\parallel}^{2d}(R)$ account for the self-interaction inside very small scroll rings. For symmetry reasons this interaction is equivalent to a boundary induced drift with the distance $z$ replaced with the filament radius $R$. The minus sign in front of $c_{\parallel}^{2d}(R)$ is due to the opposite axis orientation in relation to the spiral chirality. 

Numerical solutions of the extended kinematical model (red lines in figure \ref{fig:numtrajs}) agree quantitatively with the filament trajectories obtained from the 3d reaction-diffusion simulations. 
The dynamics of a free scroll ring with life time $R_0^2/(2\alpha)$ results if the contribution due to filament curvature is dominating over the boundary effects. This is the case when the filament radius is small, or if the distance between filament plane and no-flux boundary leads to comparably small $c_{\parallel}^{2d}(z)$ values, figures \ref{fig:exresults},\ref{fig:numtrajs}(a) and figure \ref{fig:exresultsr2} (filled circles).
%The regimes of boundary-affected filament dynamics are located in a relatively narrow range of distances around $z_{rep}$.
% 
For sufficiently large filament radii even expanding scroll rings are found in the simulation, as they have been observed in the BZ experiment, figures \ref{fig:exsetup}, \ref{fig:numtrajs}(e) and \ref{fig:exresultsr2} (open squares). 
Below initial distances $z_0 < z_{rep}$ the scroll ring is pushed by the repellor (dashed line in figure \ref{fig:numdriftfield}b) towards the no-flux boundary where it is annihilated before complete collapse. This scenario explains the experimentally observed ring collision with the boundary, figures \ref{fig:exsetup}, \ref{fig:numtrajs}(d) and \ref{fig:exresultsr2} (open circles).
%For sufficiently large filament radii even expanding scroll rings are found in the simulation, as they have been in the experiment, figures \ref{fig:exsetup}, \ref{fig:numtrajs}(e) and figure \ref{fig:exresultsr2} (open squares).
%
%Below initial distances $z_0 \leq z_{rep}$ the scroll ring is pushed by the repellor (dashed line in figure \ref{fig:numdriftfield}b) into the no-flux boundary where it is annihilated before complete collapse. 
%This scenario eplains the experimentally observed ring collision, figure \ref{fig:exsetup},\ref{fig:numtrajs}(d), figure \ref{fig:exresultsr2} (open circles).
%
For small initial ring radii within $R_0<0.6\,\lambda$, we found that taking into account the interaction between the spiral wave fronts forming the scroll ring improves the agreement between kinematical and numerical results considerably. On the kinematical level, this effect is described by the terms $c_{\parallel}^{2d}(R)$ and $c_{\perp}^{2d}(R)$ in (3), as for self-interaction inside the ring the filament radius $R$ plays the role of the distance $z$. These two terms are responsible for the drift of the ring along its symmetry axis; they become negligible small for larger core radii.
We emphasize that the \emph{transient} ring dynamics in the drift velocity field is crucial in understanding the experimental findings, particularly at the distances near $z_{rep}$. 
%\newpage
%Asymptotically for large times the influence of the no-flux boundary becomes negligible, because of the vanishingly small values of $c_{\parallel}^{2d}(z_{att})$.
%To facilitate comparisons between numerical and experimental results, space and time are scaled in units of wavelength $\lambda$ and rotation period $T$ of an unperturbed two-dimensional spiral wave, respectively.
%
%Even though all the rings considered here asymptotically vanish, we find that the interaction with the no-flux boundary induces behavior previously associated only with negative filament tension $\alpha$.
%is much richer

\begin{figure}[!ht]
\flushright
\includegraphics[width=1\linewidth]{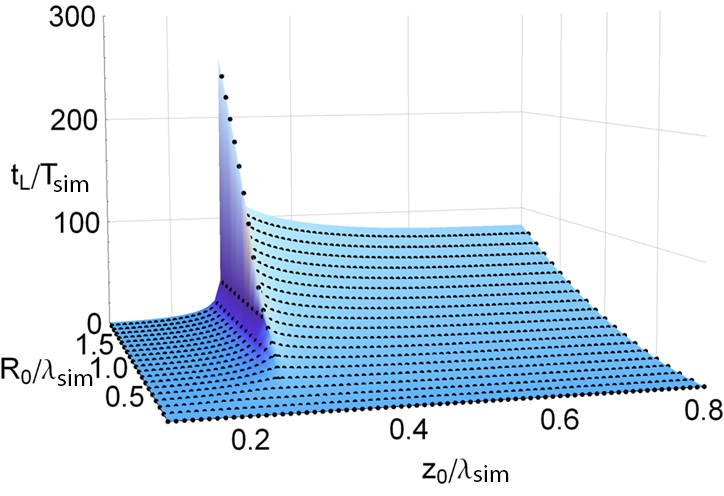}
\caption{\label{fig:lt}
%Life times of scroll rings from solutions of the modified kinematical model. In contrast to free rings, quasi-stationary rings exist with a three-fold increased lifetime. Below a critical distance, which corresponds to the repellor from two-dimensional calculations, the scroll ring lifetime is greatly reduced. Black dots are a guide for the eye.
Lifetime of spatially confined scroll rings depending on their initial radius $R_0$ and initial distance $z_0$ of their filament plane from a confining no-flux boundary. There is a marked lifetime enhancement for small scroll rings formed at distances close to the repellor of the velocity field of boundary-induced drift $z_{rep} = 0.28\,\lambda_{sim}$. Black dots are a guide for the eye. 
}
\end{figure}

From the numerical data obtained within the extended kinematical model (3), we have determined the lifetime of confined scroll rings for different initial sizes $R_0$ and distances $z_0$ (figure \ref{fig:lt}). Obviously, there is a pronounced lifetime enhancement
%boundary effect on filament dynamics 
in a range of distances close to $z_{rep}$. Here, contracting scroll rings evolve with a considerably increased lifetime, figures \ref{fig:exsetup},\ref{fig:numtrajs}(c), and \ref{fig:exresultsr2} (filled triangles), including almost persistent ones that contract at nearly vanishing rate, figures \ref{fig:exsetup},\ref{fig:numtrajs}(d) and \ref{fig:exresultsr2} (filled squares). In the ladder case, a large scroll ring initially grows and simultaneously departs from the no-flux boundary. This movement in turn weakens the interaction with the boundary and finally causes the ring to contract.

Our results are in line with numerical simulations on interacting mirror-image-like pairs of scroll rings carried out by Bray and Wikswo \cite{bray2003interaction} who observed, for example, a lifetime enhancement of the bound states. This is not surprising, as the interaction within pairs of scroll rings is for symmetry reasons equivalent to the interaction of one single scroll ring with a nearby planar no-flux boundary. For the same reason, the kinematic model elaborated in our paper is applicable to interacting pairs of scroll rings, too.

%The components of the drift velocity field are difficult to measure experimentally, especially within the interaction range of the scroll ring and the no-flux boundary. Therefore, we have determined the dependencies $c_{\parallel}^{2d}(z)$ and $c_{\perp}^{2d}(z)$ in two-dimensional numerical simulations (see figure \ref{fig:numdriftfield}a) with 

%---------------------------------------------------------- fig8 ---------------------------------------------%
%------------------------------------------------------- end fig8 ---------------------------------------------%

%

\section{Summary \& Conclusion} \label{fin}

Taking the ferroin-catalyzed BZ reaction as a representative example, we demonstrate experimentally that axisymmetric scroll rings confined to a thin layer display a rich variety of ring dynamics in excitable media with positive filament tension $\alpha$. Besides the well-known collapse of the free scroll ring of radius $R_0$ within finite time $R_0^2/2\alpha$, scroll rings interacting with a planar no-flux boundary exhibit lifetime enhancement up to at least a factor 3. These long living scroll rings act as autonomous 3d pacemakers that last over about four hours during which they change their size by less than $3.5\,\%$ (figure \ref{fig:exresults}d). Moreover, despite of positive filament tension we even found expanding scroll rings.

We have determined the evolution of the filament depending on the initial size of the ring and the distance from a planar no-flux boundary by full 3d numerical integration of the Aliev-Rovinsky reaction-diffusion model for the ferroin-catalyzed BZ reaction. In these numerical simulations we reproduced all experimentally observed regimes of ring evolution. 
%The scenarios of ring evolution in figure \ref{fig:exresults}a,b,c confirm previous numerical simulations by Bray et al. \cite{bray2003interaction} for the first time experimentally. 

We argue that the impact of spatial confinement on scroll ring evolution can be understood within a kinematic approach that combines the intrinsic dynamics of the unconfined scroll ring with boundary-induced drift of spiral waves forming the ring. To check this idea, we propose an extended kinematical model for the dynamics of the confined scroll ring that includes the distance dependent longitudinal and transversal components of the drift velocity field. The quantitative agreement between our kinematical model with the simulations validates our hypothesis that the observed phenomena are explained by a superposition of the intrinsic dynamics of the free scroll ring and boundary-induced spiral drift.

Recently, we observed stationary boundary-stabilized scroll rings in experiments with the photosensitive BZ reaction \cite{Azhand2014} as predicted long ago by Winfree \cite{winfree1989} for negative line tension. Also in this case, the experimental findings are correctly described by the extended kinematical model.

The understanding of possible confinement effects on the dynamics of scroll waves is essential for the correct interpretation of experimental data obtained in thin excitable media of the BZ reaction, in cell cultures of cardiac tissue, and many others. Regarding cardiac arrhythmias, for example, it is expected that the dynamics of transmural scroll waves \cite{cherry_visualization_2008,Biktasheva2015} in the comparatively thin atrial tissue will be affected by the interaction with the tissue boundaries. Potentially, overlooked long living transient states can play an important role as quasi-stationary pacemakers, especially when the layer thickness is comparable to the intrinsic length scale of emerging spatio-temporal patterns.

\ack

J.F.T. and H.E. thank the German Science Foundation (DFG) for financial support through the Research Training Group 1558. O.St. acknowledges support by the National Science Foundation under Grant No. 1213259.

%%%%%%%%%%%%%%%%%%%%%%%%% Bibliography %%%%%%%%%%%%%%%%%%%%%%%%%%%%%%%%%%%%%%%%%%%%%%%%%%%%%%%%%
\section*{References}
\bibliographystyle{iopart-num_jft2014}
\bibliography{bibliography}% Produces the bibliography via BibTeX.

\providecommand{\newblock}{}
\begin{thebibliography}{10}
\expandafter\ifx\csname url\endcsname\relax
  \def\url#1{{\tt #1}}\fi
\expandafter\ifx\csname urlprefix\endcsname\relax\def\urlprefix{URL }\fi
\providecommand{\eprint}[2][]{\url{#2}}

\bibitem{martens_entropic_2011}
Martens S, Schmid G, Schimansky-Geier L and H\"anggi P 2011 {\em Phys. Rev.\/}
  E \href{http://dx.doi.org/10.1103/PhysRevE.83.051135}{{\bf 83} 051135}

\bibitem{martens_hydrodynamically_2013}
Martens S, Straube A~V, Schmid G, Schimansky-Geier L and H\"anggi P 2013 {\em
  Phys. Rev. Lett.\/}
  \href{http://dx.doi.org/10.1103/PhysRevLett.110.010601}{{\bf 110} 010601}

\bibitem{Keyser2014-1}
Dettmer S~L, {Keyser} U~F and {Pagliara} S 2014 {\em Rev. {Sci}. {Instrum}.\/}
  \href{http://dx.doi.org/10.1063/1.4865552}{{\bf 85} 023708}

\bibitem{Keyser2014-2}
Dettmer S~L, {Pagliara} S, {Misiunas} K and {Keyser} U~F 2014 {\em Phys.
  {Rev}.\/} E \href{http://dx.doi.org/10.1103/PhysRevE.89.062305}{{\bf 89}
  062305}

\bibitem{gelb_phase_1999}
Gelb L~D, Gubbins K~E, Radhakrishnan R and Sliwinska-Bartkowiak M 1999 {\em
  Rep. Prog. Phys.\/} \href{http://dx.doi.org/10.1088/0034-4885/62/12/201}{{\bf
  62} 1573}

\bibitem{rayner1991vortex}
Rayner J, Thomas A, Rayner J and Thomas A 1991 {\em Phil. Trans. R. Soc.
  Lond.\/} B \href{http://dx.doi.org/10.1098/rstb.1991.0100}{{\bf 334} 107}

\bibitem{doligalski1994vortex}
Doligalski T, Smith C and Walker J 1994 {\em Ann. Rev. Fluid Mech.\/}
  \href{http://dx.doi.org/10.1146/annurev.fl.26.010194.003041}{{\bf 26} 573}

\bibitem{agladze2000waves}
Agladze K and Steinbock O 2000 {\em J. Phys. Chem.\/}
  \href{http://dx.doi.org/10.1021/jp002237n}{{\bf 104} 9816}

\bibitem{bansagi2006nucleation}
B{\'a}ns{\'a}gi T and Steinbock O 2006 {\em Phys. Rev. Lett.\/}
  \href{http://dx.doi.org/10.1103/PhysRevLett.97.198301}{{\bf 97} 198301}

\bibitem{kastberger2008social}
Kastberger G, Schmelzer E and Kranner I 2008 {\em PLoS One\/}
  \href{http://dx.doi.org/10.1371/journal.pone.0036736}{{\bf 3} e3141}

\bibitem{steinbock1993three}
Steinbock O, Siegert F, M{\"u}ller S and Weijer C 1993 {\em Proc. Natl. Acad.
  Sci.\/} \href{http://dx.doi.org/10.1073/pnas.90.15.7332}{{\bf 90} 7332}

\bibitem{zykov2011selection}
Zykov V~S, Oikawa N and Bodenschatz E 2011 {\em Phys. Rev. Lett.\/}
  \href{http://dx.doi.org/10.1103/PhysRevLett.107.254101}{{\bf 107} 254101}

\bibitem{cherry2011effects}
Cherry E and Fenton F 2011 {\em J. Theo. Biol.\/}
  \href{http://dx.doi.org/10.1016/j.jtbi.2011.06.039}{{\bf 285} 164}

\bibitem{Hartmann1996}
Hartmann N, B\"ar M, Kevrekidis I~G, Krischer K and Imbihl R 1996 {\em Phys.
  Rev. Lett.\/} \href{http://dx.doi.org/10.1103/PhysRevLett.76.1384}{{\bf 76}
  1384}

\bibitem{yang2006investigation}
Yang J and Zhang M 2006 {\em Phys. Lett.\/} A
  \href{http://dx.doi.org/10.1016/j.physleta.2005.10.065}{{\bf 352} 69}

\bibitem{brandtstadter2000experimental}
Brandtst{\"a}dter H, Braune M, Schebesch I and Engel H 2000 {\em Chem. Phys.
  Lett.\/} \href{http://dx.doi.org/10.1016/S0009-2614(00)00486-3}{{\bf 323}
  145}

\bibitem{schebesch1990}
Schebesch I and Engel H 1999 {\em Phys. Rev.\/} E
  \href{http://dx.doi.org/10.1103/PhysRevE.60.6429}{{\bf 60} 6429}

\bibitem{zemlin2005}
Zemlin C, Mukund K, Wellner M, Zaritsky R and Pertsov A 2005 {\em Phys. Rev.
  Lett.\/} \href{http://dx.doi.org/10.1103/PhysRevLett.95.098302}{{\bf 95}
  098302}

\bibitem{Winfree1973}
Winfree A~T 1973 {\em Science\/}
  \href{http://dx.doi.org/10.1126/science.181.4103.937}{{\bf 181} 937--939}

\bibitem{biktashev1994tension}
Biktashev V, Holden A and Zhang H 1994 {\em Phil. Trans. R. Soc.\/} A
  \href{http://dx.doi.org/10.1098/rsta.1994.0070}{{\bf 347} 611}

\bibitem{keener1992dynamics}
Keener J and Tyson J 1992 {\em SIAM Rev.\/}
  \href{http://dx.doi.org/10.1137/1034001}{{\bf 34} 1}

\bibitem{Panfilov1986}
Panfilov A~V, Rudenko A~N and Krinsky V~I 1986 {\em Biofizika\/} {\bf 31} 850

\bibitem{panfilov1987two}
Panfilov A and Rudenko A 1987 {\em Physica\/} D
  \href{http://dx.doi.org/10.1016/0167-2789(87)90132-1}{{\bf 28} 215}

\bibitem{bansagi2007negative}
B\'ans\'agi T and Steinbock O 2007 {\em Phys. Rev.\/} E
  \href{http://dx.doi.org/10.1103/PhysRevE.76.045202}{{\bf 76} 045202}

\bibitem{Azhand2014}
Azhand A, Totz J~F and Engel H 2014 {\em {EPL}\/}
  \href{http://dx.doi.org/10.1209/0295-5075/108/10004}{{\bf 108} 10004}

\bibitem{Storb2003}
Storb U, Neto C~R, B{\"a}r M and M{\"u}ller S~C 2003 {\em Phys. Chem. Chem.
  Phys.\/} \href{http://dx.doi.org/10.1039/B301790G}{{\bf 5} 2344--2353}

\bibitem{Alonso2003}
Alonso S, Sagu{\'e}s F and Mikhailov A~S 2003 {\em Science\/}
  \href{http://dx.doi.org/10.1126/science.1080207}{{\bf 299} 1722--1725}

\bibitem{Alonso2006}
Alonso S, Sagu{\'e}s F and Mikhailov A~S 2006 {\em J. Phys. Chem. A\/}
  \href{http://dx.doi.org/10.1021/jp064155q}{{\bf 110} 12063--12071}

\bibitem{Luengviriya2008}
Luengviriya C, {Storb} U, {Lindner} G, {M}{\"u}ller S~C, {B}{\"a}r M and
  {Hauser} M~J~B 2008 {\em Phys. {Rev}. {Lett}.\/}
  \href{http://dx.doi.org/10.1103/PhysRevLett.100.148302}{{\bf 100} 148302}

\bibitem{Dahmlow2013}
D{\"a}hmlow P, Alonso S, B{\"a}r M and Hauser M~J~B 2013 {\em Phys. Rev.
  Lett.\/} \href{http://dx.doi.org/10.1103/PhysRevLett.110.234102}{{\bf 110}
  234102}

\bibitem{Linde1991}
Linde H and {Engel} H 1991 {\em Physica\/} D
  \href{http://dx.doi.org/10.1016/0167-2789(91)90188-F}{{\bf 49} 13--20}

\bibitem{amemiya_formation_1998}
Amemiya T, Kettunen P, K{\'a}d{\'a}r S, Yamaguchi T and Showalter K 1998 {\em
  Chaos\/} \href{http://dx.doi.org/10.1063/1.166373}{{\bf 8} 872}

\bibitem{Jimenez2012}
Jim\'{e}nez Z~A and Steinbock O 2012 {\em Phys. Rev.\/} E
  \href{http://dx.doi.org/10.1103/PhysRevE.86.036205}{{\bf 86} 036205}

\bibitem{Rov1984}
Rovinskii A~B and Zhabotinskii A~M 1984 {\em J. Phys. Chem.\/}
  \href{http://dx.doi.org/10.1021/j150669a001}{{\bf 88} 6081}

\bibitem{Aliev1992}
Aliev R and Rovinskii A 1992 {\em J. Phys. Chem.\/}
  \href{http://dx.doi.org/10.1021/j100181a039}{{\bf 96} 732}

\bibitem{taylor_scroll_1999}
Taylor A~F, Johnson B~R and Scott S~K 1999 {\em Phys. Chem. Chem. Phys.\/}
  \href{http://dx.doi.org/10.1039/a809339c}{{\bf 1} 807--811}

\bibitem{Muller1987}
M{\"u}ller S~C, Plesser T and Hess B 1987 {\em Physica\/} D
  \href{http://dx.doi.org/10.1016/0167-2789(87)90068-6}{{\bf 24} 87--96}

\bibitem{pertsov_three-dimensional_1993}
Pertsov A, Vinson M and M{\"u}ller S~C 1993 {\em Physica\/} D
  \href{http://dx.doi.org/10.1016/0167-2789(93)90157-V}{{\bf 63} 233--240}

\bibitem{Field1972}
Field R~J, {Koros} E and {Noyes} R~M 1972 {\em {J}. {Am}. {Chem}. {Soc}.\/}
  \href{http://dx.doi.org/10.1021/ja00780a001}{{\bf 94} 8649--8664}

\bibitem{bray2003interaction}
Bray M~A and Wikswo J~P 2003 {\em Phys. Rev. Lett.\/}
  \href{http://dx.doi.org/10.1103/PhysRevLett.90.238303}{{\bf 90} 238303}

\bibitem{Ermakova1989}
Ermakova E~A, {Pertsov} A~M and {Shnol} E~E 1989 {\em Physica\/} D
  \href{http://dx.doi.org/10.1016/0167-2789(89)90062-6}{{\bf 40} 185--195}

\bibitem{Gomez-Gesteira1996b}
{G}{\'o}mez {Gesteira} M, {Mu}{\~n}uzuri A~P, {P}{\'e}rez {Mu}{\~n}uzuri V and
  {P}{\'e}rez {Villar} V 1996 {\em Phys. {Rev}.\/} E
  \href{http://dx.doi.org/10.1103/PhysRevE.53.5480}{{\bf 53} 5480--5483}

\bibitem{Biktasheva2010}
Biktasheva I~V, Barkley D, Biktashev V~N and Foulkes a~J 2010 {\em Phys.
  Rev.\/} E \href{http://dx.doi.org/10.1103/PhysRevE.81.066202}{{\bf 81}
  066202}

\bibitem{Aranson1991}
Aranson I~S, Kramer L and Weber A 1991 {\em Physica\/} D
  \href{http://dx.doi.org/10.1016/0167-2789(91)90069-L}{{\bf 53} 376}

\bibitem{Aranson1993}
Aranson I~S, {Kramer} L and {Weber} A 1993 {\em Phys. {Rev}.\/} E
  \href{http://dx.doi.org/10.1103/PhysRevE.47.3231}{{\bf 47} 3231--3241}

\bibitem{winfree1989}
Nandapurkar P~J and Winfree A~T 1989 {\em Physica\/} D
  \href{http://dx.doi.org/10.1016/0167-2789(89)90070-5}{{\bf 35} 277}

\bibitem{cherry_visualization_2008}
Cherry E~M and Fenton F~H 2008 {\em New J. Phys.\/}
  \href{http://dx.doi.org/10.1088/1367-2630/10/12/125016}{{\bf 10} 125016}

\bibitem{Biktasheva2015}
Biktasheva I, Dierckx H and Biktashev V 2015 {\em Phys. Rev. Lett.\/}
  \href{http://dx.doi.org/10.1103/PhysRevLett.114.068302}{{\bf 114} 068302}

\end{thebibliography}

\end{document}